\documentclass{iaus}
\usepackage{graphicx,psfig}

\title[LMXBs in Sculptor] 
{The discovery of X-ray binaries in the Sculptor Dwarf Spheroidal Galaxy}

\author[short author list]   
{Thomas J. Maccarone$^1$, Arunav Kundu$^2$, Stephen E. Zepf$^2$, Anthony L. Piro$^3$ \and Lars Bildsten$^{3,4}$}

\affiliation{$^1$School of Physics and Astronomy,University of Southampton, Southampton, SO17 1BJ, UK \break email:tjm@phys.soton.ac.uk\\[\affilskip]
$^2$Department of Physiscs and Astronomy, Michigan State University, East Lansing, MI, USA\\[\affilskip]
$^3$Department of Physics, University of California at Santa Barbara, Santa Barbara, CA, USA\\[\affilskip]
$^4$Kavli Institute for Theoretical Physics,University of California at Santa Barbara, Santa Barbara, CA, USA\\}

\pubyear{2005}
\volume{230}  
\pagerange{}
\date{}
\jname{Populations of High Energy Sources in Galaxies}
\editors{E. Meurs \& G. Fabbiano, eds.}
\begin{document}

\maketitle

\begin{abstract}
We report the results of a deep Chandra survey of the Sculptor dwarf
spheroidal galaxy.  We find five X-ray sources with $L_X$ of at least
$6\times10^{33}$ ergs/sec with optical counterparts establishing them as
members of Sculptor.  These X-ray luminosities indicate that these
sources are X-ray binaries, as no other known class of Galactic point
sources can reach 0.5-8 keV luminosities this high.  Finding these
systems proves definitively that such objects can exist in an old
stellar population without stellar collisions.  Three of these objects
have highly evolved optical counterparts (giants or horizontal branch
stars), as do three other sources whose X-ray luminosities are in the
range which includes both quiescent low mass X-ray binaries and the
brightest magnetic cataclysmic variables.  We predict that large area
surveys of the Milky Way should also turn up large numbers of
quiescent X-ray binaries.
\end{abstract}

\firstsection 
\section{Introduction}
It is quite difficult for field star populations to produce low mass
X-ray binaries (LMXBs), especially those with neutron star primaries.
Supernova explosions which eject more than half the mass of a system
normally leave the systems unbound.  Therefore the only ways to
produce LMXBs through binary stellar evolution are through common
envelope evolution which ejects much of the mass of the black hole's
or neutron star's progenitor before the supernova occurs (Paczynski
1976; Kalogera \& Webbink 1998), or through a finely-tuned asymmetric
velocity kick which occurs at the birth of the neutron star or black
hole (Brandt \& Podsiadlowski 1995; Kalogera 1998).  Intermediate mass
X-ray binaries (i.e. X-ray binaries where the donor star is 1-8
$M_\odot$) can lose large amounts of mass and then evolve into LMXBs,
which may help to solve the problem of formation rates of such LMXBs,
since these systems will be more likely to survive the supernova
explosions (see e.g. Podsiadlowski, Rappaport \& Pfahl 2002). Because
of the uncertainties in how to keep LMXBs bound, and other
uncertainties regarding, e.g. the correlation between the initial
masses of the stars in a binary system, theoretical rates of X-ray
binary formation are highly uncertain.

The difficulties in producing LMXBs are even larger for old stellar
populations.  The mass accretion rate required to power a persistent
bright LMXB is about $10^{-8}M_\odot$yr$^{-1}$, meaning that the
lifetime of such a system can be only about 100 Myrs from when
accretion starts.  It is generally believed that most LMXBs will begin
their accretion phases only a few Gyrs after the supernova which
creates the compact object (White \& Ghosh 1998).  Several
possibilities still exist for producing X-ray binaries in old
populations such as elliptical galaxies -- these LMXBs we see in old
stellar populations may be very low duty cycle transients (meaning
that they will take a long time to accrete their entire mass donors)
which begin accretion only after their donor stars evolve off the main
sequence (Piro \& Bildsten 2002), they may be ultracompact X-ray
binaries (Bildsten \& Deloye 2004), or they may be normal X-ray
binaries which were produced through dynamical encounters in globular
clusters, and then released into the field (e.g. White, Sarazin \&
Kulkarni 2002; see also Grindlay 1988).

If the bulk of the elliptical galaxy field sources are low duty cycle
transients, then there must exist an underlying population of
quiescent X-ray binaries which is much larger than the fraction which
is flaring at any given time.  Therefore, we have identified a large
sample of old stars which is nearby, and does not contain any globular
clusters, the Sculptor dwarf spheroidal galaxy.  By finding five X-ray
binaries in this galaxy, four of which have highly evolved donor
stars, we have verified that the mechanism of Piro \& Bildsten (2002)
is, at the very least, a substantial contributor to the X-ray binary
populations seen in old field star populations.

\section{Data}
We have obtained 21 exposures of 6-kiloseconds each with the Chandra
X-ray observatory over the time period from 26 April 2004 to 10
January 2005.  The monitoring, rather than a single deep observation,
was performed in order to search for bright transient sources, but
none were found.  We then stacked the data to make a single image of
the data on ACIS-S3, and ran WAVDETECT, finding 74 sources in the
0.5-8.0 keV band.  Based on previous deep field measurements
(e.g. Hornschemeier et al. 2001), we estimate that about 50 of these
sources should be background AGN and that it is unlikely that more
than one is a foreground star. In order to determine whether there are
supersoft X-ray sources, we have also extracted an image from 0.1-0.4
keV, but this image contained no bright sources (i.e. with more than
10 photons) that were not found in the 0.5-8.0 keV image.

We have compared the positions of the 9 sources with at least 100
detected counts (giving accurate positions, and making them bright
enough to rule out the possibility of a white dwarf accretor) with the
positions of bright optical stars ($V<20.5$) in the Sculptor galaxy
from Schweitzer et al. (1995).  The stars in the Schweitzer et
al. (1995) catalog are all red giants, asymptotic giant branch stars,
or horizontal branch stars.  We have also checked whether the single
brightest source has an optical counterpart in the deep optical
photometry of Hurley-Keller, Mateo \& Grebel (1999).

Allowing for an 0.6'' boresight correction we find four matches within
0.4'' of an optical star in Schewitzer et al. (1995).  The chance
superposition probability is 0.04 for these objects, and all have
proper motion confirmations that they are members of Sculptor, rather
than foreground or background sources.  The fifth match is the
brightest X-ray source in our sample, which is found at RA=1h00m13.9s,
Dec=-33d44m42.5s.  While there is no strong evidence for it to be
variable in our monitoring campaign, it was not detected in the ROSAT
bright source all-sky survey, so it must either be variable at the
level of a factor of a few, or it must have a very hard spectrum at
soft X-rays. Its optical counterpart is $R=23.68,B-R=0.95$
(D. Hurley-Keller private communication - see Hurley-Keller et
al. 1999 for a description of how the photometry was done), giving it
a luminosity and color roughly consistent with a solar-type star in
the low metallicity environment of the Sculptor galaxy, and it appears
to be near the turnoff of the main sequence in the Sculptor galaxy.

\section{Discussion}
\subsection{The nature of the matches}
It has been suggested that the field populations of elliptical
galaxies are low duty cycle transients, with red giant donors (Piro \&
Bildsten 2002).  Obviously the finding that several of these systems
have red giant donors provides support for this hypothesis.  The
number of bright (i.e. $L_X>10^{37}$ ergs/sec) LMXBs per unit stellar
mass seems to be roughly constant across a sample of giant elliptical
galaxies (Gilfanov 2004), with a typical value of about one per
$10^8-10^9 M_\odot$ of stars.  We would then expect about $\sim$ .02
such systems in Sculptor, given its mass of about 2$\times10^6
M_\odot$ (Mateo 1998).  Given that the duty cycles of these transients
are likely to be less than about 1/200 in the scenario of Piro \&
Bildsten (2002), one would expect that the number of quiescent
transient X-ray binaries in the Sculptor galaxy would be at least a
few, and possibly much higher if the duty cycles were even smaller
than 1/200.  Our discovery of one clear case of an X-ray binary with a
red giant counterpart and a few more candidates is thus in reasonable
agreement with the picture suggested by Piro \& Bildsten (2002).

\subsection{Are there other quiescent X-ray binaries in Sculptor?}
It is likely that there are actually many more quiescent X-ray
binaries in the Sculptor galaxy than the five strong matches we have
presented here and the three additional tenative matches.  Our
detection limits in this data set are at a luminosity of a little bit
above $10^{32}$ ergs/sec, while quiescent LMXBs have been seen to be
as faint as $2\times10^{30}$ ergs/sec, with most of the faintest
sources having black hole accretors (Garcia et al. 2001).  Therefore,
some quiescent X-ray binaries could be below our sensitivity limits,
although most neutron star accretors are brighter than these limits in
quiescence, and there do seem to be higher quiescent luminosities for
long period transients, even for black hole accretors.  There may also
be other quiescent X-ray binaries which are detected in our
observations, but which have faint optical counterparts.  These could
be, for example, ultracompact X-ray binaries, which have also been
suggested to be systems which should be present in old stellar
populations (Bildsten \& Deloye 2004), or even X-ray binaries with
lower main sequence donors which simply took longer than the typical
few Gyrs (White \& Ghosh 1998) to come into contact.  We plan to
investigate these possibilities in future work with deeper optical
photometry and with spectroscopic follow-ups.

\subsection{Implications for X-ray binaries in other stellar populations}
These results clearly indicate that the canonical factor of 100
enhancement in X-ray binaries per unit stellar mass in globular
clusters applies only to bright LMXBs.  Comparing this galaxy with
NGC~6440, which shows the highest number density of quiescent LMXBs
(Heinke et al. 2003), we find that the globular cluster density is
enhanced by a factor of only about 10, even assuming that we have
found all the quiescent LMXBs in Sculptor already.  Since other
clusters are less rich in LMXBs than NGC~6440, the enhancement factor
is even lower on the whole.  This is probably because very wide
binaries such as those with red giant donors do not survive
unperturbed for a Hubble time in dense globular clusters.

Finally, we consider the implications of these results for the Milky
Way's bulge's X-ray binary population.  The bulge of the Milky Way
contains about $5\times10^2$ times as many stars as the Sculptor dwarf
spheroidal galaxy, meaning that it should have about 2,500 X-ray
binaries if the number of binaries scales linearly with mass.  In
fact, since X-ray binary production is enhanced in metal rich systems
(at least for globular clusters - see Kundu, Maccarone \& Zepf 2003
for evidence of this effect and Maccarone, Kundu \& Zepf 2004 for a
discussion of a binary evolution model for this effect which should
work equally well in field populations as in globular clusters) the
Milky Way might contain a few times more than this number of X-ray
binaries.  Our estimate is comparable to numbers from theoretical
predictions (Iben et al. 1997; Belczynski \& Taam 2004).

Finally, we note that with only about 500 optical stars in the
Schweitzer et al. (1995) catalog, at least $\approx$1\% of the giant
branch/horizontal branch stars in Sculptor are in systems with
accreting neutron stars or black holes.  Based on this finding, one
would expect that a large number of similar systems are likely to be
found in surveys that include hundreds of red giant stars.  This
hypothesis should be testable as the results from the ChaMPlane survey
(Grindlay et al. 2003) begin to come in, or in a large area survey of
the Galactic Bulge.

\begin{acknowledgments}
We are grateful to Carine Babusiaux, Kyle Cudworth, Peter Jonker,
Christian Knigge, Mike Muno, Eric Pfahl, Andrea Schweitzer, Jeno
Sokoloski, Kyle Westfall and Rudy Wijnands for useful discussions.
We thank Denise Hurley-Keller, Eva Grebel and Mario Mateo for the
optical photometric measurement of the brightest X-ray source.  AK and
SEZ acknowledge support from NASA grants SAO G04-5091X and LTSA
NAG-12975.  LB acknowledges support from the NSF via grant PHY99-07949
\end{acknowledgments}

\end{document}